\documentclass[showpacs,aps,pra,
twocolumn]{revtex4-1}
\usepackage{graphicx}
\usepackage{amsfonts}
\usepackage{amsmath}
\usepackage{colordvi}
\usepackage{color}

\usepackage{epsfig}

\begin{document}
\title{Linear and nonlinear  Zeno effects in an optical coupler}
\author{F. Kh. Abdullaev$^{1}$,  V. V. Konotop$^{1,2}$,   and V. S. Shchesnovich$^{3}$}
\affiliation{ $^1$Centro de F\'isica Te\'orica e Computacional, Faculdade de Ci\^encias,
Universidade de Lisboa,   Avenida Professor
Gama Pinto 2, Lisboa 1649-003, Portugal
$^2$Departamento de F\'isica, Faculdade de Ci\^encias,
Universidade de Lisboa, Campo Grande, Ed. C8, Piso 6, Lisboa
1749-016, Portugal
\\
$^3$Centro de Ci\^encias Naturais e Humanas, Universidade Federal do ABC, Santo Andr\'e, SP 09210-170, Brazil
}

\begin{abstract}
It is shown  that in a simple coupler where one of the  waveguides is subject to
controlled   losses of the electric field, it is possible to observe optical analog
of the linear and nonlinear quantum Zeno effects. The phenomenon consists in a
\textit{counter-intuitive}  enhancement of transparency of the coupler with the
increase of the dissipation and represents an optical analogue  of the quantum Zeno
effect. Experimental realization of the phenomenon based on the use of the
chalcogenide glasses is proposed. The system allows for observation of the
cross-over between the linear and nonlinear Zeno effects, as well as effective
manipulation of  light transmission  through the coupler.
\end{abstract}
\maketitle

\newpage

\section{Introduction}

Decay of a quantum system,  either because it is in a  metastable state or due to
its interaction with an external system (say, with a measuring apparatus), is one
of the fundamental problems of the quantum mechanics. Already more  than fifty
years ago it was proven that the decay of a quantum metastable system is, in
general,  non-exponential~\cite{Khalfin,Degasperis} (see also the
reviews~\cite{rev,Khalfin_rev}). Ten years later in Ref.~\cite{Misra} it was
pointed  that a quantum system undergoing   frequent  measurements does not decay
at all in the limit of the infinitely-frequent measurements. This remarkable
phenomenon was termed by the authors  the quantum ``Zeno's paradox''. The Zeno's
paradox, i.e. the total inhibition of the decay,  requires, however,  unrealistic
conditions and shows only as the Zeno effect, i.e. the decrease of the decay rate
by frequent observations, either pulsed or continuous. The Zeno effect was observed
experimentally by studying the decay of continuously counted beryllium
ions~\cite{beryl}, the escape of cold atoms from an accelerating optical
lattice~\cite{zeno_lat}, the control of spin motion by circularly polarized
light~\cite{zeno_spin}, the decay of the externally driven mixture of two hyperfine
states of rubidium atoms~\cite{zeno_BEC}, and the production of cold molecular
gases~\cite{zeno_molec}. There is also the opposite effect, i.e. the acceleration
of the decay by observations, termed the anti-Zeno effect, which is even  more
ubiquitous in the quantum systems  \cite{Zeno}.

It was  argued that the quantum Zeno and anti-Zeno effects can be explained from
the purely dynamical point of view, without any reference to the projection
postulate of the quantum mechanics \cite{FPJPhyA}. In this respect, in
Ref.~\cite{BKPO,SK} it is shown that the Zeno effect can   be understood within the
framework of the mean field description, when the latter can be applied,  thus
providing the link between purely quantum and classical systems.

The importance of the Zeno effect goes beyond the quantum systems. An   analogy
between the quantum Zeno effect and the decay of light in an array of optical
waveguides, was suggested in Ref.~\cite{Longhi}. Namely, the authors found an exact
solution which showed a non-exponential decay of the field in one of the
waveguides. Modeling of the quantum Zeno effect in the limit of frequent
measurements using down conversion of light in the sliced nonlinear crystal was
considered in Ref. \cite{Reh}. The effect has been mimicked by the wave process in
a $\chi^{(2)}$ coupler with a linear and nonlinear arms, since in the strong
coupling limit the pump photons propagate in the nonlinear arm without decay. The
analogy between the inhibition of losses of molecules and the enhanced reflection
of light from a medium with a very high absorption was also noticed
in~\cite{zeno_BEC}.

Meantime, in the mean field models explored in Refs.~\cite{BKPO,SK} inter-atomic
interactions play an important role, leading to the nonlinear terms in the
resulting dynamical equations. In its turn, the nonlinearity  introduces
qualitative differences in the Zeno effect, in particular dramatically reducing the
decay rate~\cite{SK}   compared to the case of noninteracting atoms. This
phenomenon,  the enhancement of the   effect by the inter-atomic interactions, in
Ref.~\cite{SK} was termed the  {\em nonlinear Zeno effect} (since, when  the
nonlinearity is negligible, it reduces to the usual linear Zeno effect).

Mathematically, the mean field description of a Bose-Einstein condensate (BEC) and
of the light propagation in Kerr-type media are known to have many similarities,
due to the  same (Gross-Pitaevskii or nonlinear Schr\"odinger) equation describing
the both phenomena. {Furthermore, the linear Zeno effect, is observable not only in
pure quantum statement, but also in the meanfield approximation~\cite{SK}. This
immediately suggests that detecting  the Zeno dynamics, is possible in the
classical systems, and in particular, in the nonlinear optics, thus offering new
possibilities for managing of light~\cite{com1}. Namely,  one can expect the
counter-intuitive  reduction of   attenuation of the total field amplitude (which
would correspond to the reduction of   losses  of   atoms in the BEC case) by
increasing the losses in some parts of the system (an analogy to the increasing of
the  removal rate of  atoms in the case of BEC). }

To report on a very basic system where analogs of the  linear and nonlinear Zeno  effects can
be observed and exploited is the main goal of the present paper. More
specifically, we explore the mathematical analogy  of the semi-classical dynamics of a  BEC in a
double well potential subject to removal of atoms~\cite{SK}, with light propagation
in a nonlinear optical coupler, in which one of the arms is subject to controllable
losses.

The paper is organized as follows. First in Sec.~\ref{sec:two_examp}  we consider
two well known models of dissipative oscillators, which illustrate the classical
analogues of the Zeno phenomenon (originally introduced in the quantum measurement
theory). Next, in Sec.~\ref{sec:experiment} we discuss possible experimental
settings allowing observation of the phenomenon in optics. In
Sec.~\ref{sec:NonLinZeno} the theory of optical nonlinear Zeno effect is considered
in details. Sec.~\ref{sec:lin_nonlin} is devoted to comparative analysis of the
linear and nonlinear Zeno effects. The outcomes are summarized in the Conclusion.

\section{Two trivial examples.}
\label{sec:two_examp}

Before going into details of the optical system, let us first give a
simple insight on the pure classical origin of the phenomenon of inhibition of the field attenuation by
strong dissipation. First, we recall the well-known fact, that increase of the
dissipation $\alpha$ of an overdamped  ($\alpha\gg \omega$) oscillator $\ddot{x} +
\alpha \dot{x}+\omega^2 x=0$, results in decrease of the attenuation of the oscillations. Indeed the decay rate $R\approx
\omega^2/\alpha$ approaches zero, when the dissipation coefficient $\alpha$ goes to
infinity. But the amplitude of oscillations in this case is also nearly zero.
However, the coupling of another linear oscillator to the dissipative one,
\begin{eqnarray*}
\ddot{x}_1+\alpha\dot{x}_1+\omega^2 x_1+\kappa x_2=0,
\quad
\ddot{x}_2+ \omega^2
x_2+\kappa x_1=0,
\end{eqnarray*}
allows one to observe the inhibition of attenuation due to strong
dissipation by following a finite amplitude $x_2$. Indeed, the characteristic equation,
$$
\displaystyle{\lambda=\frac{\lambda^4+2\lambda^2\omega^2+\kappa^2-\omega^4}{\alpha(\lambda^2+\omega^2)}},
$$
evidently has the small root
$
\lambda\approx (\kappa^2-\omega^4)/ (\alpha\omega^2)
$
which appears for $\alpha\gg \kappa^2/\omega^2-\omega^2>0$. Thus, one of the
dynamical regimes of the system is characterized by the decay rate which goes to
zero in the overdamped case, moreover the relation between the amplitudes of the
damped and undamped oscillators reads  $|x_1/x_2|\to \omega^2/\kappa<1$ as
$\alpha\to\infty$. In other words, strong dissipation in one of the
oscillators can attenuate  the energy decay in the whole system. On the other hand,
the last example illustrates that if the coupling is of the same order as the
eigenfrequencies of the subsystems, the energy is distributed between the two
subsystems in approximately equal parts. This does not allow for further decrease
of the decay rate of the energy, because its large part is concentrated in the
damped subsystem.

{The phenomenon descried above for the linear oscillators can be viewed as a
classical analog of the linear Zeno effect.} The nonlinearity changes the situation
dramatically. This case, however does  not allow for complete analytical treatment,
any more, and that is why we now turn to the specific nonlinear system, which will
be studies numerically. We consider an optical coupler composed of two Kerr-type
waveguides,  one arm of which is subject to relatively strong field losses. We will
show that such a coupler mimics the quantum Zeno effect,  allowing one to follow,
in a simple optical experiment, the cross-over between the linear {(weak
intensities)} and nonlinear {(strong intensities)} Zeno effects, thus providing a
deep analogy between the effect of dissipation in the classical and quantum
systems. In particular, we will also show that  {\em strong losses of the field in
one of the waveguides can significantly enhance the transmittance of the coupler as
a whole}.

\section{The coupler  and a possible experimental setting.}
\label{sec:experiment}

The optical fields in
the two tunnel-coupled nonlinear  optical fibers~\cite{Jensen82} (alternatively,
one can consider two linearly couple waveguides~\cite{planar}) are described by the
system
\begin{subequations}
\label{sys}
\begin{eqnarray}
\label{sys_a}
-i\frac{da_1}{dz} =(\beta_1 +i\alpha_1)a_1 \pm \gamma |a_1|^2 a_1 + \kappa a_2,\\
\label{sys_b}
-i\frac{da_2}{dz} =(\beta_2 +i\alpha_2)a_2 \pm \gamma |a_2|^2 a_2 + \kappa a_1.
\end{eqnarray}
\end{subequations}
Here $a_{1,2}$ are the properly normalized fields in each arm of the  coupler,
$\kappa$ is the coupling coefficient measuring the spatial overlap between the
channels, the upper and lower signs correspond to the focusing ($+$) and defocusing
($-$) media, $\beta_j $ ($j=1,2$) are the modal propagation constants of the cores,
$\gamma = 2\pi n_2/\lambda A_{eff}$, $n_2$ is the Kerr nonlinearity parameter,
$A_{eff}$ is the effective cross section of the fiber, $\lambda$ is the wavelength,
and the loss coefficient $\alpha_{j} >0$ stands for the field absorption in the $j$th waveguide.

Our aim is to employ the manageable losses, i.e. a control over the
coefficients $\alpha_{1,2}$, in order to observe different regimes of the light
transmission through the coupler. Since in optics one cannot easily manipulate with
$z$, i.e. with the length of the coupler, we are interested in realizing different
dynamical regimes with a single given coupler (rather than using  several couplers
having different characteristics). This contrasts with the BEC case where the
propagation variable $z$ corresponds to time (see e.g.~\cite{SK}) and can be easily
varied. For this reason the most suitable experimental setting could be with a
coupler, whose properties strongly depend on the wavelength of the input beam
(alternatively one can consider flexible change of the optical properties using
temporal gradients, active doping, etc.).

An experimentally feasible realization of the described nonlinear directional
coupler can be based on the use of the $As_2 Se_3$ chalcogenide glass. For this
material the intrinsic nonlinearity can be up to three orders of magnitude greater
than that of the pure silica fibers~\cite{Taeed,Chremmos,Ruan}.  More specifically,
one can consider the material losses in the chalcogenide glasses, where the Kerr
nonlinearity parameter is $n_2 =1.1 \cdot 10^{-13}$cm$^2$/W, that is 400 times of
the nonlinearity of the fused silica fiber. However, what is even more important
for our aims, is that the absorption rate of at least one of the coupler arms can
be changed dramatically during the experiment. Say, in the mentioned chalcogenide
glasse $\alpha$  can be on the order of a few dB/m and is very sensitive to the
wavelength. Thus, a practical control over the absorption can be performed by
using the dependence of the loss coefficients $\alpha_{1,2}$ on the
wavelength of the incident light.

To implement this idea, it is necessary to produce the two arms of the coupler
using chalcogenide  glasses of different types. In particular, one can consider the
standard sulphide fiber in one arm of the coupler  and the lowest loss sulphide
fibers~\cite{Aggarwal} in the other arm. Such sulphide fibers have a particularly
narrow attenuation peak at  the wavelength  $\lambda_0\approx 3\mu$m. The behavior
of the absorption coefficient $\alpha$  in the vicinity of this peak can be modeled
by the Lorentzian curve (here we use the experimental results reported
in~\cite{Loren}):
\begin{eqnarray}
\label{Lorenz}
\alpha_{1}(\lambda) = \alpha_{1,0} + \frac{\alpha_{1,1} \Gamma^2}{(\lambda- \lambda_0)^2 + \Gamma^2},
\end{eqnarray}
where $\Gamma \approx (0.5\div 1)\,\mu$m and $\alpha_{1,0} \approx 0.5$dB/m and
$\alpha_{1,1} \sim 5$dB/m for a usual sulfide fiber.
 Varying the wavelength  about $\lambda_0\approx 3\mu$m in   the interval
$\lambda_0\pm 0.5\mu$m the loss can be varied in the standard sulfide fiber by
$(0.5\div 5)$dB/m, and in the lowest  loss sulfide fiber in the interval $(0.05\div
0.2)$dB/m.  Even larger  attenuation  can be achieved for chalcogenide fibers
30Ge-10As-30Se-30Te,  where the pick attenuation is  on the order of $(5 \div
30)dB/m$ observed for $\lambda \approx 4.5\mu$m.

\section{The nonlinear optical Zeno effect.}
\label{sec:NonLinZeno}

Thus we  consider the situation when one of the waveguides (waveguide 1) is subjected
to controllable  losses (as discussed above), while the second one
(waveguide 2) is operating in the transparency regime, i.e. when $\alpha_1 \gg
\alpha_2$. Respectively, we simplify the problem setting in what  follows
$\alpha_2=0$.

We start with the estimate of the effective losses, designated below as
$\widetilde{\alpha}_2$, in the transparent arm of the coupler, which occur due to
the energy exchange between the arms. For $z\gg \alpha_1^{-1}$, one can
adiabatically eliminate $a_1$ from the system (\ref{sys}). Moreover, assuming that
$\alpha_1 \gg \kappa$ we obtain: $|a_1|^2 \approx
\frac{\widetilde{\alpha}_2}{\alpha_1}|a_2|^2\ll|a_2|^2$ and
\begin{equation}
-i\frac{d a_2}{dz} \approx \left(i\widetilde{\alpha}_2 + \widetilde{\beta}_2 +
\widetilde{\gamma}|a_2|^2\right)a_2.
\label{EQa2}
\end{equation}
Here  $\widetilde{\beta}_2 = \beta_2 +
 \widetilde{\alpha}_2 (\beta_2-\beta_1)/\alpha_1$  and  $\widetilde{\gamma}
= \gamma\left(1+ \widetilde{\alpha}_2/\alpha_1\right)$ ,   with the effective
$z$-dependent attenuation rate:
\begin{equation}
\widetilde{\alpha}_2 = \frac{\alpha_1\kappa^2}{\left( \beta_2-\beta_1  +
 \gamma |a_2|^2\right)^2 +  \alpha_1^2 }.
\label{newrate}
\end{equation}
First of all, we  observe that $\widetilde{\alpha}_2$ decays  with increase of the
difference $\beta_2-\beta_1$ or the nonlinearity (the term $\gamma|a_2|^2$ in the
denominator). This behavior is natural because the difference in the propagation
constants $\beta_{1,2}$ results in incomplete energy transfer between the arms,
whereas the nonlinearity effectively  acts as an additional amplitude-dependent
detuning. In practical terms, however, the effect due to the constant linear
detuning is negligible, because $\beta_2-\beta_1$ is typically too small, whereas
the nonlinearity can result in an appreciable effect. Thus, the effective
attenuation rate $\widetilde{\alpha}_2$ decays either with increase of the
absorption $\alpha_1\to\infty$ (the linear Zeno effect), or (for given losses
$\alpha_1$) with the intensity of the light in the transparent  arm, tending to
zero in the formal limit $ |a_2|^2\to \infty$ (the nonlinear Zeno effect).

Moreover, the anti-Zeno effect, i.e. the increase of the attenuation with the
increase of the loss coefficient $\alpha_1$ can be observed merely due to the
presence of a strong nonlinearity. Such an  effect, however, is not
counter-intuitive  in our setup. In fact, it is rather trivial:   for $\gamma
|a_2|^2\gg \alpha_1$ Eq. (\ref{newrate}) tells us that  the ratio $|a_1|^2/|a_2|^2
\approx \widetilde{\alpha}_2/\alpha_1$ is independent of $\alpha_1$, which means
that, in the Zeno regime $z\gg \alpha_1^{-1}$, increasing the loss coefficient must
increase the actual attenuation.

In order to perform the complete numerical study of  the coupler
 we introduce the real amplitudes and phase of the fields:    $a_j = \rho_j
\exp(i\phi_j)$, $j=1,2$, the relative difference in the energy flows in the two
arms  $F = (|a_1|^2 - |a_2|^2)/(|a_1|^2 + |a_2|^2)$,   the total
energy flow in the coupler $Q = (|a_1|^2 + |a_2|^2)/P_0$, normalized to the input
flow $P_0 = |a_{10}|^2 + |a_{20}|^2$, as well as the phase mismatch $\phi = \phi_1
- \phi_2$.
 Then
the original system~(\ref{sys}) is reduced to
\begin{subequations}
\label{sys1}
\begin{eqnarray}
\label{sys1_a}
&&F_Z = -g (1-F^2) + 2\sqrt{1-F^2}\sin(\phi), \\
\label{sys1_b}
&&\phi_Z = \frac{(\beta_1 -\beta_2)}{\kappa} \pm 4\delta F Q  - 2\frac{F}{\sqrt{1-F^2}}\cos(\phi),\\
\label{sys1_c}
&&Q_Z = - g Q(1+F)
\end{eqnarray}
\end{subequations}
where $g=\alpha_1/\kappa$, $\delta = P_0/P_c$, and the distance is normalized on
the linear coupling length $L = 1/\kappa$, i.e. $Z =z/L = z\kappa$. Here we also
introduced the critical power $P_c = 4\kappa/\gamma$, The critical power $P_c$
which  separates the regimes with periodic exchange energy between the arms for $P
< P_c$ and the localization of energy in one of the waveguides if $P >
P_c$~\cite{Jensen82}. For a fiber based on the chalcogenide glass, described above,
the critical power $P_c\sim 1$W and the coupling length L varies in the interval
($0.1 \div 1$)m.

Notice that mathematically the system~(\ref{sys1}) coincides with the  one
describing a BEC in a double-well trap subject to elimination of atoms from one of
the wells~\cite{SK}. The coupler mimics the nonlinear Zeno effect in a BEC in a
double-well trap, where the time is replaced by the propagation distance in the
coupler  and the electric fields in the arms of the coupler correspond to the
number of quantum particles in the potential wells. The system (\ref{sys1}) also
resembles the evolution of Bose-Hubbard dimer with non-Hermitian Hamiltonian~\cite{Korch}

\section{Linear {\it vs} nonlinear Zeno effects.}
\label{sec:lin_nonlin}

Passing to the numerical study of the system (\ref{sys1}) we estimate that for the
length $L$ on the order of one meter the value of the absorption coefficient $g$
can be changed in chalcogenide fibers by up to $20$ times.  In the empiric formula
(\ref{Lorenz}) the values of the dimensionless parameters are as follows:
$g_{1,0}=\alpha_{1,0}/\kappa \sim 1$ and $g_{1,1} \alpha_{1,1}/\kappa \sim (10\div
20)$, while $\delta = P_0/P_c$ is in the interval $(0 \div 2.5)$.

\begin{figure}[f]
\epsfig{file=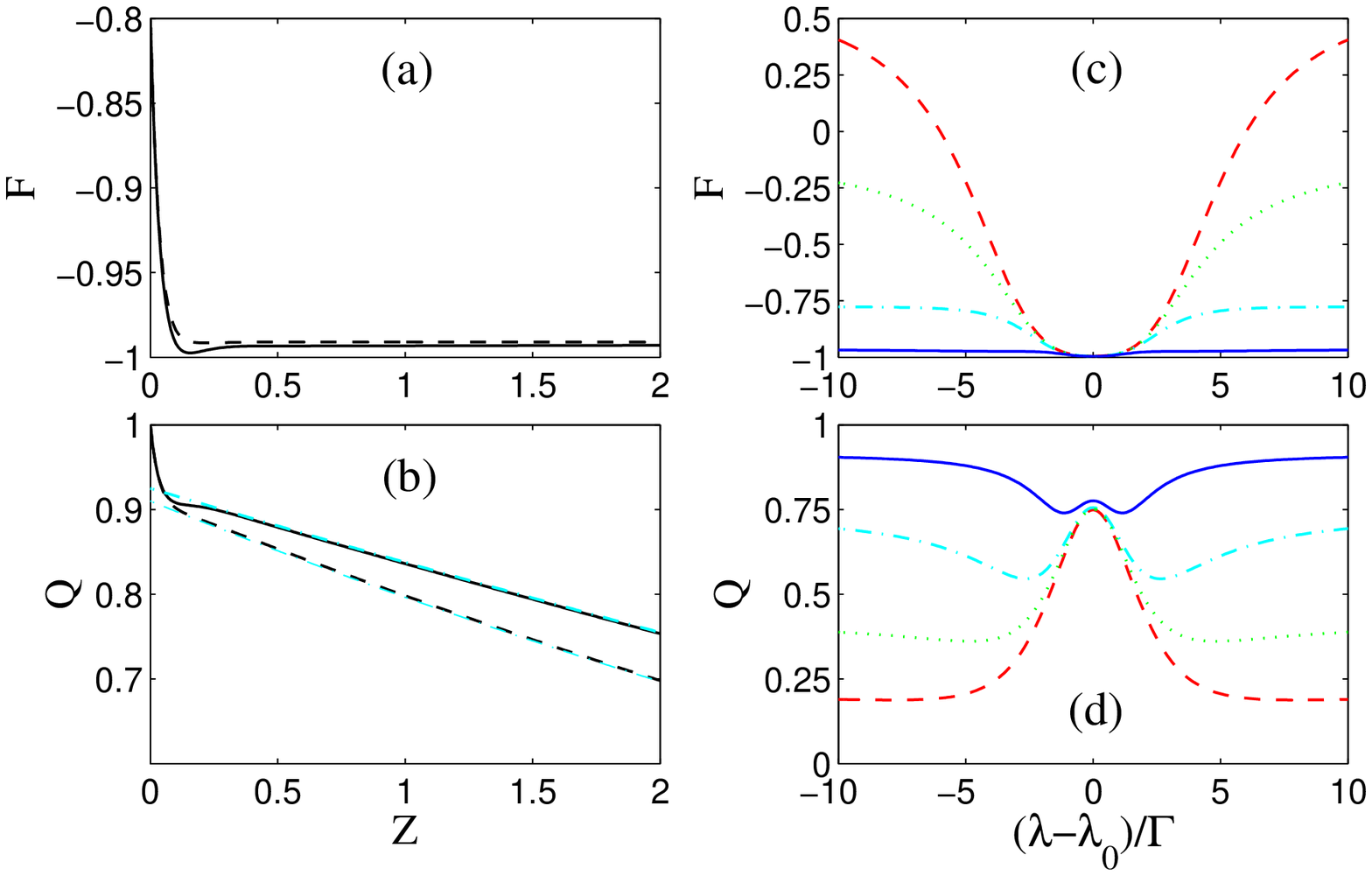,width=\columnwidth}
\caption{ (Color online) The relative
(a), and the total (b)  energy flows {\it vs} the propagation distance $Z$. Here
$F(0) = -0.8$, $\phi(0) = 0$, $\beta_2=\beta_1=0$, $g = 15$, $\delta = 0$ (dashed
lines) and $\delta = 2.5$ (solid lines). The dash-dotted lines in panel (b) show
the reduced rate attenuation given by Eq.~(\ref{EQa2}) (also obtained by
numerical   simulations).  In panels (c) and (d): the output energy
distribution ($F$, panel (c)) and the total output energy flow ($Q$, panel (d)) in
the coupler of the length $Z = 2$, as functions of the  deviation of the input
light wavelength from the attenuation pick intensity $\lambda_0$. The output
results for different values of $\delta$, $\delta = 0$ (dashed), $\delta = 0.5$
(dots), $\delta = 1$ (dash-dot) and $\delta = 2.5$ (solid), demonstrate the linear
and nonlinear Zeno effects.}
\label{FG1}
\end{figure}

Our main results are summarized in Fig.~\ref{FG1}. In panels (a) and (b), where we
show the dependence of the output signal {\it vs} the coupler length,  the three
different regimes are evident. At small distances from the coupler input,
$Z\lesssim 0.2$, the standard {\em exponential} decay occurs. This stage does not
depend significantly on the intensity of the input pulse (i.e. on $\delta$ in our
notations). However at larger distances,  $0.2\lesssim Z\lesssim 2$, the system
clearly reveals {\em power-like decay}. The power of the decay, however appears to
be sensitive to the magnitude of the input power, i.e. to the nonlinearity of the
system. The decay is  much stronger at lower powers ($\delta\approx0$),
corresponding to the  linear  Zeno effect, and much weaker  for the input
intensities above the critical value ($\delta=2$), and  may be termed  the {\em
nonlinear} Zeno effect. In all the cases the output beam is concentrated in the
fiber/waveguide without losses [Fig.~\ref{FG1}a] and the output power is still
sufficiently high above 70\% of the imput power.  We also notice that, while we
have chosen relatively large $g$ the phenomenon is also observable (although less
pronounced) for lower levels of the light absorption.

In practice, however, Figs~\ref{FG1} (a) and (b) will not correspond to a real
experiment with an optical coupler, because in the standard settings its length,
i.e. $Z$, is fixed. Instead, as we mentioned above, observation of the Zeno
effects,  can be achieved by  varying wavelength of the light.  From the panel (b)
one concludes that the best observation of the phenomenon can be achieved at some
intermediate lengths of the coupler, where, on the one hand, the power-like decay
is already established and, on the other hand, the output power is still high so
that the system is still in the nonlinear regime (we do not show transition to the
linear regime, which for the data used in Fig.~\ref{FG1} occurs approximately at
$Z\approx 0.3$. In our case the coupler lengths satisfying the above requirements
correspond to the interval $0.2\lesssim Z\lesssim 2$. Respectively, choosing $Z=2$
in the panels (c) and (d) we show how the output intensity depends on the
wavelength of the incident beam, which can be manipulated experimentally. In,
particular in panel (d) one clearly observes the linear Zeno effect as a dramatic
increase of the output power (the transparency window of the coupler) exactly at
the pick attenuation (see the dashed curve) achieved at the wavelength $\lambda_0$,
as well as practically lossless propagation of the field in the nonlinear
case (c.f. the solid line with dashed curves). Remarkably, for the strongly
nonlinear case we also observe local increase of the output power, which however is
preceded by the small decay of the power. The local decay of the intensity appears
when the input power is approximately equal to the critical one ($\delta\approx
1$).

\begin{figure}[f]
\begin{center}
\epsfig{file=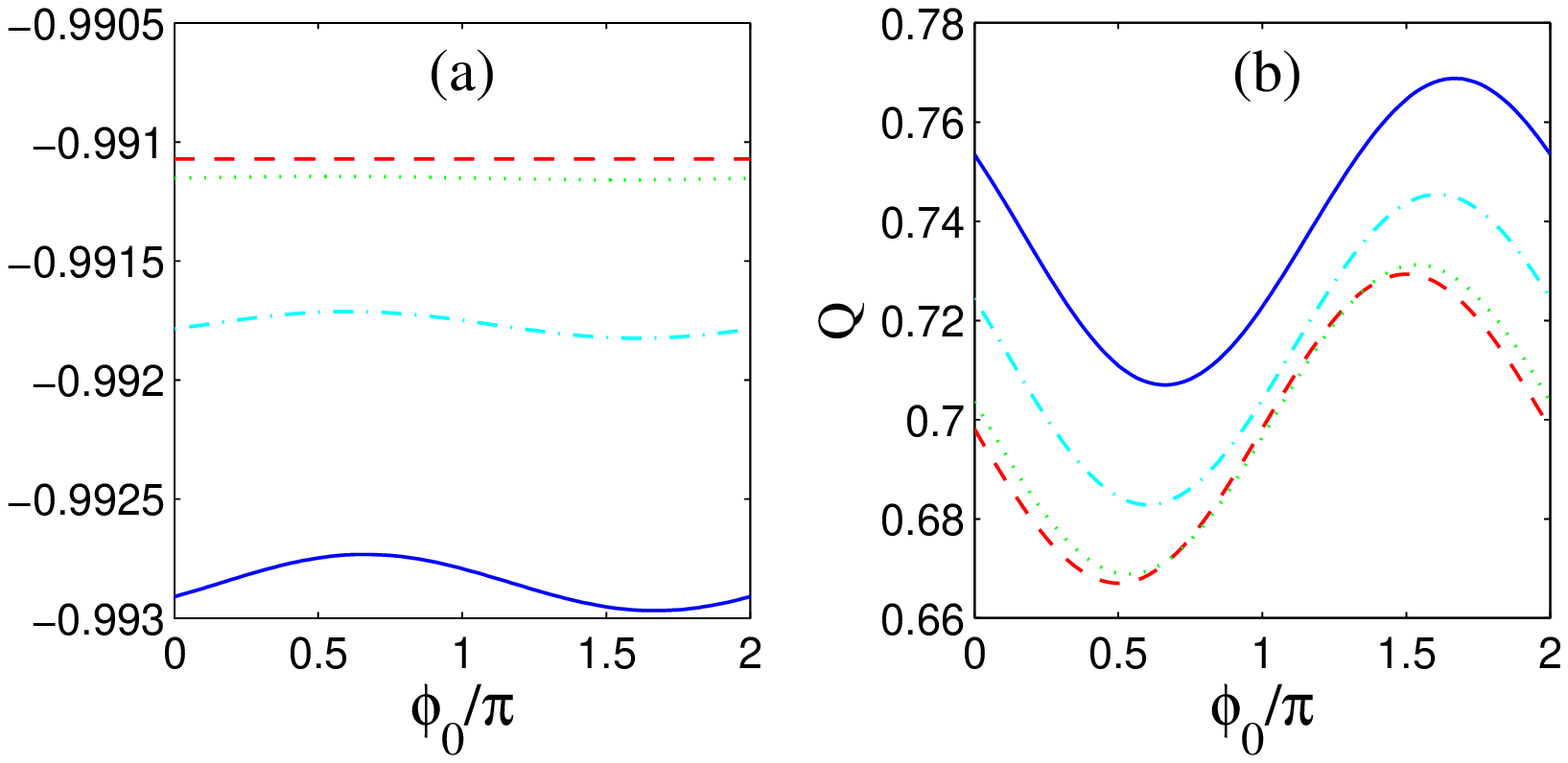,width=\columnwidth} \caption{(Color online) Output energy
distribution ($F$,  panel (a)) and the total output energy ($Q$, panel (b)) {\it
vs} the relative input phase $\phi(0)$. Here $F(0) = -0.8$, $g = 15$, $Z = 2$ and
$\beta_2=\beta_1=0$. The lines correspond to different values of $\delta$,
representing the nonlinear Zeno effect: $\delta = 0$ (dashed), $\delta = 0.5$
(dots), $\delta = 1.5$ (dash-dot) and $\delta = 2.5$ (solid).}
\label{FG2}
\end{center}
\end{figure}

So far, however, we have considered the case of zero phase mismatch between the two
arms of the coupler. In fig. \ref{FG2} we show the dependence on the phase mismatch
between the two cores.

One observes that the input phase mismatch does not destroy the phenomenon, but can
affect the output energy flow by the order of 10\% (the relative energy
distribution being practically unchanged).

\section{Conclusions}

To conclude, we have shown that by using a simple optical coupler subjected to the
wavelength  dependent absorption of the light in one of the arms, one can observe
the linear and nonlinear Zeno effects. The phenomenon consists in increase of the
output energy with increase of the   absorption coefficient  of  one of the arms.
The linear Zeno effect shows an especially strong dependence on the wavelength of
the input signal, {as this is expected from the design of the system}. The
nonlinear Zeno effect, observed at the intensities above the critical one, is
characterized by a much larger transparency of the system, {and consequently}
accompanied by much weaker  dependence on the input wavelength.

It is interesting to mention that, recently, the effect of a light  localization in
linear coupler with the strong losses in one waveguide has been
observed~\cite{DOC1}. The authors attributed this phenomenon to the {\cal
PT}-symmetric configuration of their passive coupler (to which it can be reduced by
the proper change of variables). Since the presence of the nonlinearity rules out
the referred change of variables, the present work  proposes an alternative
explanation of the experiment reported in~\cite{DOC1}, and moreover, shows that
this is a quite general phenomenon (non necessarily related to the {\cal
PT}-symmetry), which can be observed in linear and nonlinear systems and which open
new possibilities for manipulating transmission of light by means of controllable
absorption by  making it either intensity of wavelength dependent.

\acknowledgments

The authors greatly acknowledge stimulating discussions with Alex Yulin. FKA and
VVK were  supported by  the 7th European Community Framework Programme under the
grant PIIF-GA-2009-236099 (NOMATOS). VSS was supported by the CNPq of Brasil.


\begin{thebibliography}{99}
\bibitem{com1} Notice, that the optical analogy of the linear quantum Zeno effect described
in~\cite{Longhi} was based on existence of evanescent modes, i.e. on a  different physical
principle having and had very different physical manifestation.
\end{thebibliography}
\end{document}